\documentclass{article}
\usepackage{spconf,amsmath,graphicx,amssymb}
\usepackage{textcomp,siunitx,multicol}
\usepackage[utf8]{inputenc}
\usepackage{verbatim,balance}
\usepackage{amsthm,lipsum,cite}
\usepackage{tikz,siunitx,verbatim}
\usepackage[inline]{enumitem}

\usepackage{pgfplots}
\pgfplotsset{compat=newest}
\pgfplotsset{plot coordinates/math parser=false}
\newlength\figureheight
\newlength\figurewidth

\def\w{\mathbf w}

\def\P{\mathbf P}

\def\T{\mathbf T}

\title{Overlap-Add Windows with Maximum Energy Concentration\\ for Speech and Audio Processing}
\name{Tom Bäckström\thanks{This project was supported by the Academy of Finland research project No 312490.}}
\address{Aalto University, Department of Signal Processing and Acoustics, Espoo, Finland}
\begin{document}

\maketitle
\begin{abstract}
  Processing of speech and audio signals with time-frequency representations require
  windowing methods which allow perfect reconstruction of the original signal and where
  processing artifacts have a predictable behavior. The most common approach for this purpose
  is overlap-add windowing, where signal segments are windowed before and after processing.
  Commonly used windows include the half-sine and a Kaiser-Bessel derived window. The latter
  is an approximation of the discrete prolate spherical sequence, and thus a maximum energy concentration
  window, adapted for overlap-add. We demonstrate that performance can be improved by including the overlap-add
  structure as a constraint in optimization of the maximum energy concentration criteria.
  The same approach can be used to find further special cases such as optimal
  low-overlap windows. Our experiments demonstrate that the proposed windows provide
  notable improvements in terms of reduction in side-lobe magnitude.
\end{abstract}
\begin{keywords}
time-frequency processing, windowing, discrete prolate spherical sequences
\end{keywords}

\section{Introduction}
Speech and audio signals are slowly time-varying in character, such that it is beneficial to analyze and
process them in short segments. When the segment length is chosen appropriately, we can treat the signal
as a stationary process within the segment such that statistical modeling becomes efficient. Many applications
then use time-frequency transforms on the segments such as the short-time Fourier transform or the modified
discrete cosine transform, for the benefit of statistical and perceptual efficiency~\cite{benesty2008springer,Bosi:2003,backstrom2017celp,vilkamo2017timefrequency}.

Segmenting a signal is a windowing problem, where the segment is extracted by multiplying with a windowing
function, which is non-zero in a limited range. In analysis applications, signal processing has a long history
in the design of such windowing functions and its theory is presented in every basic book of signal
processing, e.g.~\cite{Mitra:1998}. The principal objective of windowing in analysis applications is to
minimize the detrimental effect of windowing on the signal statistics. In processing applications, however,
we also need to consider the effect of windowing on the reconstruction process.

\begin{figure}[t]
  \centering  
  \includegraphics[width=.9\columnwidth]{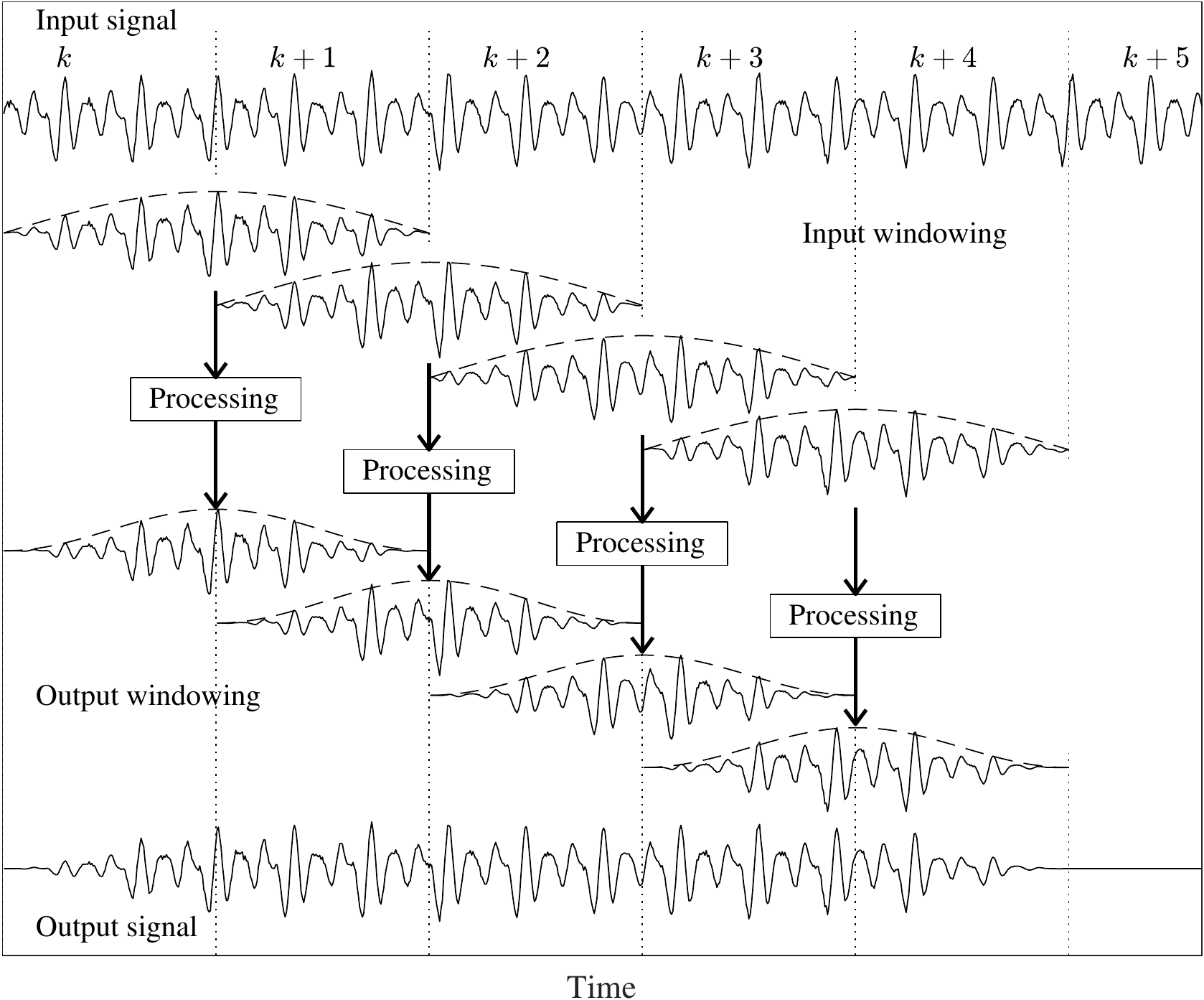}
  \caption{Illustration of input and output windowing with overlap-add synthesis in a speech processing application.}\label{fg:ola_example}
  
\end{figure}

A widely used approach in time-frequency processing of signals is known as overlap-add, where the input
signal is windowed into overlapping segments, and after processing, the segments are windowed a second
time before adding them together~\cite{harris1978use,nuttall1971spectral,vilkamo2017timefrequency} (see Fig.~\ref{fg:ola_example}).
By a careful choice of windowing functions, we can ensure that, in the absence of modifications to the
windowed signal, the original signal can be reconstructed from the windowed segments. This is known
as the \emph{perfect reconstruction} property.

Windowing is often discussed in combination with time-frequency transforms, whence the combination is
known as a filterbank~\cite{boashash2015time}. A particular type of filterbanks are those which, in
addition to perfect reconstruction, also provide critical sampling. The most commonly used critically
sampled filterbank in audio processing is the modified discrete cosine
transform~\cite{edler1989codierung,malvar1990lapped,malvar1992signal}, which can also be applied in a
bit-exact manner~\cite{geiger2001audio}. Typically, such applications use the half-sine or
a Kaiser-Bessel derived (KBD) window, which are some of the few windows applicable in overlap-add.
A radically different approach is commonly used in speech coding
with code-excited linear prediction (CELP), where temporal correlation is explicitly modeled by a
linear predictor, such that the predictor residual can be windowed without
overlaps~\cite{backstrom2013:win,backstrom2017celp}.

The performance of windows which are suitable for overlap-add have however not received the same
rigorous attention as the classical windowing methods. This paper presents methods for representing
the symmetries required by overlap-add as constraints such that the window performance can be
optimized. Specifically, we will use the maximum energy concentration
criteria~\cite{barbosa1986maximum}, familiar
from Slepian or discrete prolate spherical sequence (DPSS) -windows, to obtain optimal windows for overlap-add.

\section{Overlap-Add Windowing}
The objectives, when applying windowing in a processing applications, are two-fold:
\begin{enumerate*}
\item In the absence of any modifications, we require that the original signal can be reconstructed perfectly.
\item When the windowed signal is modified, then the energy expectation of the modification (or error),
  in the output signal, should be uniform over time.
\end{enumerate*}

Let $x_k$ be our input signal which we want to segment into overlapping windows. Window $h$
of the signal is then
\begin{equation}
    y_{k,h} = w_{k-Lh/2} x_k,
\end{equation}
where $w_k$ is the windowing function of length $L$ defined as
\begin{equation}
    \begin{cases}
       w_k > 0, & \text{when~}k\in[1,L] \\
       w_k = 0, & \text{when~}k\leq 0 \text{~or~} k> L.
    \end{cases}
\end{equation}
We can then apply some processing on the windows $y_{k,h}$ such that the modified
signal is $\hat y_{k,h}$.

To reconstruct the signal, we apply windowing again by multiplying with the windowing
function and add the windows together, such that the modified output signal is
\begin{equation}
    \hat x_k := \sum_{h} w_{k-Lh/2} \hat y_{k,h}.
\end{equation}
It is important to observe that the window is applied twice, once on the input signal
and a second time after processing on the modified output window. Only after applying the
window twice can we add the segments together to obtain the resynthesised signal.

\begin{figure}[t]
    \centering
    \resizebox{.8\columnwidth}{!}{
    \begin{tikzpicture}[scale=0.8]
     \begin{scope}[shift={(0,0)}]
     \node at (0.25,1.75) {(a)};
     \draw[ultra thick,dashed] (2,0) sin (4,3);
     \draw[ultra thick,dashed] (6,0) sin (4,3);
     \draw[ultra thick] (4,0) sin (6,3);
     \draw[ultra thick] (8,0) sin (6,3);
     \draw[ultra thick,dotted] (6,0) sin (8,3);
     \draw[ultra thick,dotted] (10,0) sin (8,3);
     \node at (4,3.5) {$h-1$};
     \node at (6,3.5) {$h$};
     \node at (8,3.5) {$h+1$};
     \node at (6,4) {Window index};
     \draw[->,very thick] (2,0) -- (10.5,0);
     \draw[->,very thick] (2,0) -- (2,3.5);
     \node[rotate=90] at (1.7,3) {$1$};
     \node[rotate=90] at (1.7,0) {$0$};
     \node[rotate=90] at (1,1.75) {Magnitude $w_k$};
     \draw (2,3) -- (2.2,3);
     \end{scope};
     
     \begin{scope}[shift={(0,-4)}]
     \node at (0.25,1.75) {(b)};
     \draw[ultra thick,dashed] (3,1.5) sin (4,3);
     \draw[ultra thick,dashed] (5,1.5) sin (4,3);
     \draw[ultra thick,dashed] (3,1.5) sin (2,0);
     \draw[ultra thick,dashed] (5,1.5) sin (6,0);
     \draw[ultra thick] (5,1.5) sin (6,3);
     \draw[ultra thick] (7,1.5) sin (6,3);
     \draw[ultra thick] (5,1.5) sin (4,0);
     \draw[ultra thick] (7,1.5) sin (8,0);
     \draw[ultra thick,dotted] (7,1.5) sin (8,3);
     \draw[ultra thick,dotted] (9,1.5) sin (8,3);
     \draw[ultra thick,dotted] (7,1.5) sin (6,0);
     \draw[ultra thick,dotted] (9,1.5) sin (10,0);
     \draw (4,3) -- (8,3);
     \node at (6,-.5) {Sample position $k$};
     \draw[->,very thick] (2,0) -- (10.5,0);
     \draw[->,very thick] (2,0) -- (2,3.5);
     \node[rotate=90] at (1.7,3) {$1$};
     \node[rotate=90] at (1.7,0) {$0$};
     \node[rotate=90] at (1,1.75) {Magnitude $w_k^2$};
     \draw (2,3) -- (2.2,3);
     \end{scope};
    \end{tikzpicture}}
    \caption{(a) Typical input windowing functions of subsequent frames and (b) the corresponding
      squared windowing functions. The thin line in (b) demonstrates the region where
    the windows add up to unity as required by Eq.~\ref{eq:princenbradley}.}
    \label{fg:princen}
\end{figure}
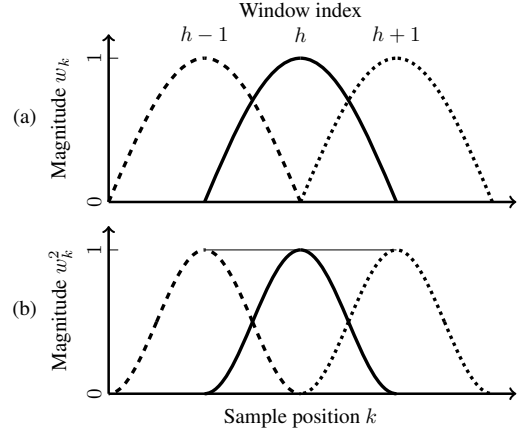

It is well-known and we can readily see that both the requirement of perfect reconstruction
and uniform error energy is ensured when the windowing function satisfies the Princen-Bradley
criteria~\cite{backstrom2017celp,Bosi:2003}
\begin{equation}\label{eq:princenbradley}
  w_{k+L/2}^2  + w_{k}^2 = 1,\qquad\text{for~}k\in[1,\,L/2].
\end{equation}
Figure~\ref{fg:princen} illustrates a typical windowing function which satisfies the Princen-Bradley
criteria and Figure~\ref{fg:ola_example} illustrates the effect of overlap-add windowing on a speech
signal.

\section{Constrained Maximization of\\ Energy Concentration}
Windowing in the time-domain corresponds to convolution in the frequency-domain. To minimize
frequency-domain distortion, we therefore require that energy of the windowing function
in the frequency-domain is maximally concentrated. The concentration of energy can be
evaluated by the ratio of energy in the pass-band versus total energy
\begin{equation}
  \tau = \frac{\int_{-\delta}^\delta |W(f)|^2 df}{\int_{-\infty}^\infty |W(f)|^2 df},
\end{equation}
where $W(f)$ is the spectrum of the windowing function and $\delta$ is the bandwidth of the pass-band.
For discrete, finite length windowing functions $\w=[w_1,\dotsc,\,w_{L}]^T$
it can be shown that the above ratio is equivalent with
\begin{equation}\label{eq:maxene}
  \tau = \frac{\w^T \T \w}{\|\w\|^2},
\end{equation}
where $\T\in{\mathbb R}^{L\times L}$ is a symmetric Toeplitz matrix with elements
\begin{equation}\label{eq:toep}
  \T_{k-h}=\frac{L \sin\left(\frac\pi L \alpha (k-h)\right)}{(k-h)},
\end{equation}
where $\alpha\in(0,1)$ defines the width of the main lobe. Clearly the maximum of $\tau$ is then the eigenvector of $\T$
corresponding the largest eigenvalue and we can equivalently define
\begin{equation}\label{eq:maxeig}
  \max \w^T \T\w\text{~such that~} \|w\|^2 = 1.
\end{equation}
The eigenvectors of $\T$ are known as discrete prolate
spherical sequences (DPSS) and the corresponding windowing functions are known correspondingly
as DPSS or Slepian windows~\cite{barbosa1986maximum,slepian1961prolate,simons2010slepian}.

The main objective of this paper is to design windowing functions which fulfills those
symmetries required by overlap-add processing, while simultaneously optimizing the above spectral
characteristics. 
The Princen-Bradley conditions of Eq.~\ref{eq:princenbradley} can then be written as
\begin{equation}\label{eq:matconstr}
  \w^T \P_k\w = 1,
\end{equation}
where $\P_k$ is diagonal with diagonal entries $[\P_k]_{h,h}= \delta_{k-h}+\delta_{k+L/2-h}$. In
other words, $\P_k$ has two non-zero entries on the diagonal which pick out the $k$th and $(k+L/2)$th
samples of the windowing vector $\w$. Consequently, the matrices $\P_k$ are positive semi-definite.
Observe that the constraints Eq.~\ref{eq:matconstr} is similar to the constraint in Eq.~\ref{eq:maxeig}
but more strict. We can therefore define a new optimization problem, using the constraints of
Eq.~\ref{eq:matconstr} and the objective function of Eq.~\ref{eq:maxeig} as
\begin{equation}\label{eq:qcqp}
    \max \w^T \T \w\text{~such that~} \w^T \P_k \w = 1\text{~for~} k\in[1,\,L/2].
\end{equation}
This is a quadratically constrained quadratic programming (QCQP) problem, which is known to be convex
if the matrices $\T$ and $\P_k$ are positive definite. We can therefore use numerical optimization
based interior-point methods to find the optimal solution.

\section{Low-overlap Windows}
\begin{figure}[t]
    \centering
    \resizebox{.8\columnwidth}{!}{
    \begin{tikzpicture}[scale=0.8]
     \begin{scope}[shift={(0,0)}]
     \draw[ultra thick,dashed] (2,0) sin (4,3) -- (6,3);
     \draw[ultra thick,dashed] (8,0) sin (6,3);
     \draw[ultra thick] (6,0) sin (8,3) -- (10,3);
     \draw[ultra thick] (12,0) sin (10,3);
     \draw[dotted] (6,3.5) -- (6,0) -- (4,0) -- (4,3.5);
     \draw[dotted] (8,3.5) -- (8,0) -- (10,0) -- (10,3.5);
     \node at (5,3.5) {$h-1$};
     \node at (9,3.5) {$h$};
     \node at (6,4) {Window index};
     \draw[->,very thick] (2,0) -- (12.5,0);
     \draw[->,very thick] (2,0) -- (2,3.5);
     \node[rotate=90] at (1.7,3) {$1$};
     \node[rotate=90] at (1.7,0) {$0$};
     \node[rotate=90] at (1,1.75) {Magnitude $w_k$};
     \draw (2,3) -- (2.2,3);
     \node at (7,-.5) {Sample position $k$};
     \end{scope};
     
    \end{tikzpicture}}
    \caption{To obtain a low overlap between windows, we can constrain a number of samples
      in the middle of the window to have unit magnitude. Thin dotted vertical lines indicate
      the borders between the flat tops and overlap areas.}
    \label{fg:low}
\end{figure}
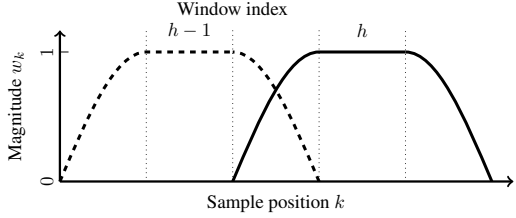

In some applications, it is desirable to limit the overlap length between
windows~\cite{fuchs2015low}. The conventional approach in designing windows of length $L$ with overlap $T$, is to
choose a windowing function of length $2T$ and extend it by a vector of $L-2T$ ones in the middle,
such that the desired length is achieved (see Fig.~\ref{fg:low}). This heuristic method can now be amended using the
optimization presented above.

Specifically, we can define new constraints as
\begin{equation}
  \left\{
  \begin{array}{rcll}
    w_{k+L-T}^2  + w_{k}^2 &=& 1,&\text{for~}k\in[1,\,T]. \\
    w_{k} & = &1,&\text{for~}k\in(T,L-T).
  \end{array}\right.
\end{equation}
Substituting these quadratic and linear constraints into the optimization problem of Eq.~\ref{eq:qcqp}
yields a low-overlap window which has maximal energy concentration.

\section{Evaluation}
The most commonly used overlap-add windows include the half-sine and a Kaiser-Bessel derived
(KBD) window. The half-sine window is defined as
\begin{equation}
  w_{\sin,k} = \sin\left(\frac{\pi\left(k-\frac12\right)}{L}\right),\text{~for~}k\in[1,L].
\end{equation}
The KBD window is based on the Kaiser-Bessel window, defined as
\begin{equation}\label{eq:kaiser}
  u_k = I_0\left[\alpha\sqrt{1-\left(\frac{2(k-\frac12)}{L-1}-1\right)^2}\right],\text{~for~}k\in[1,L],
\end{equation}
where $\alpha>0$ specifies the width of the main-lobe. The KBD window is then defined as
\begin{equation}
  w_{KBD,k} = \sqrt{\frac{\sum_{h=1}^k u_k}{\sum_{h=1}^L u_k}},\text{~for~}k\in[1,L].
\end{equation}
In other words, the KBD takes the cumulative sum of the Kaiser-Bessel window, normalizes it by the sum
and then takes a square root to satisfy Princen-Bradley.

\begin{figure}
  \centering
  \includegraphics[width=.9\columnwidth]{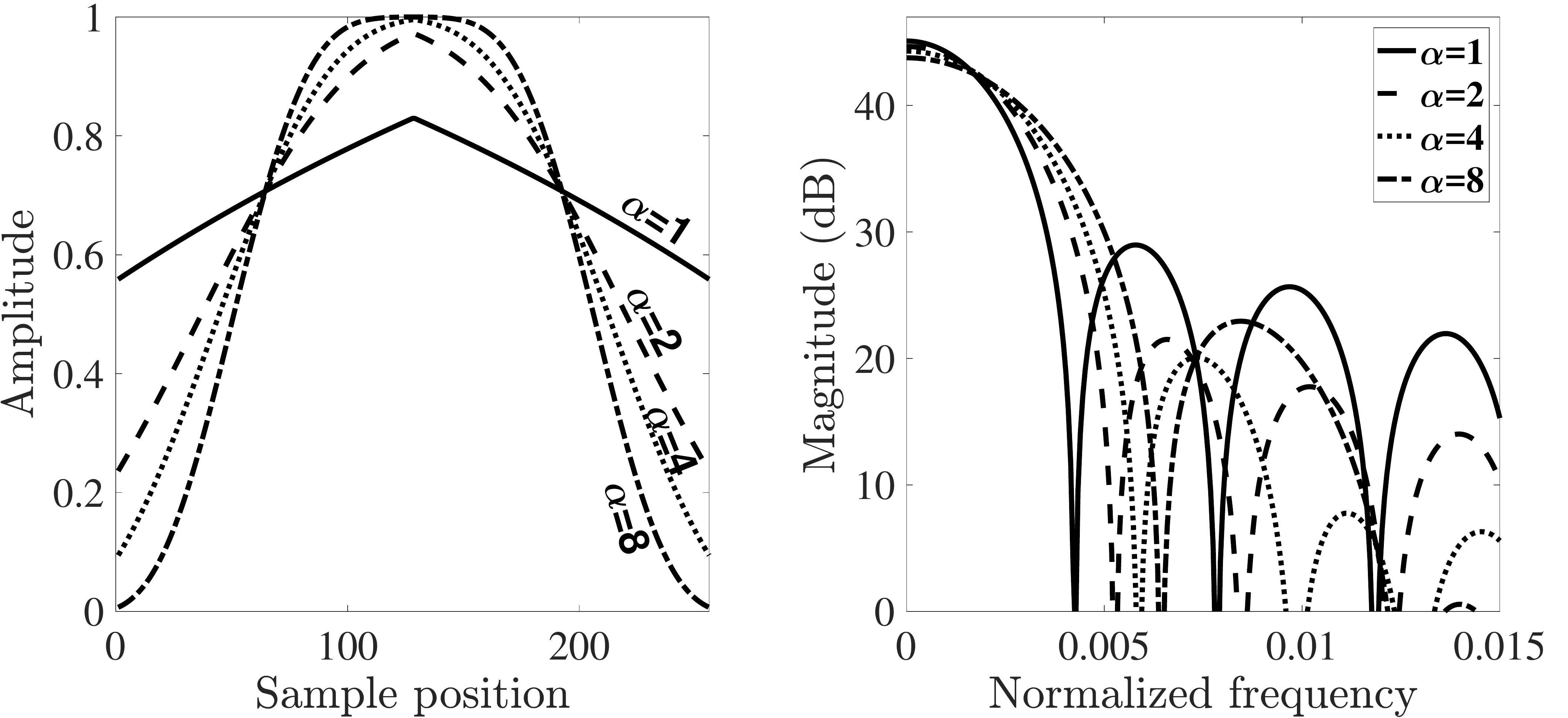}
  \caption{Illustration of the shapes of proposed window with different values of $\alpha$ and a window length of $L=256$.}\label{fg:win_shapes}
  
\end{figure}

We generated the proposed DPSS based overlap-add windows (OLA-DPSS) by using the interior-point algorithm
of the Optimization toolbox in Matlab2018a. Fig.~\ref{fg:win_shapes} demonstrates the obtained window
shapes for different values of the parameter $\alpha$.
As an informal observation, we did not have any problems with
convergence and the running times were only some seconds. Since windowing functions are usually determined
off-line, we conclude that computational capacity is not an issue in calculation of OLA-DPSS windows.

\begin{figure*}[t]
  \centering
  \includegraphics[width=0.78\textwidth]{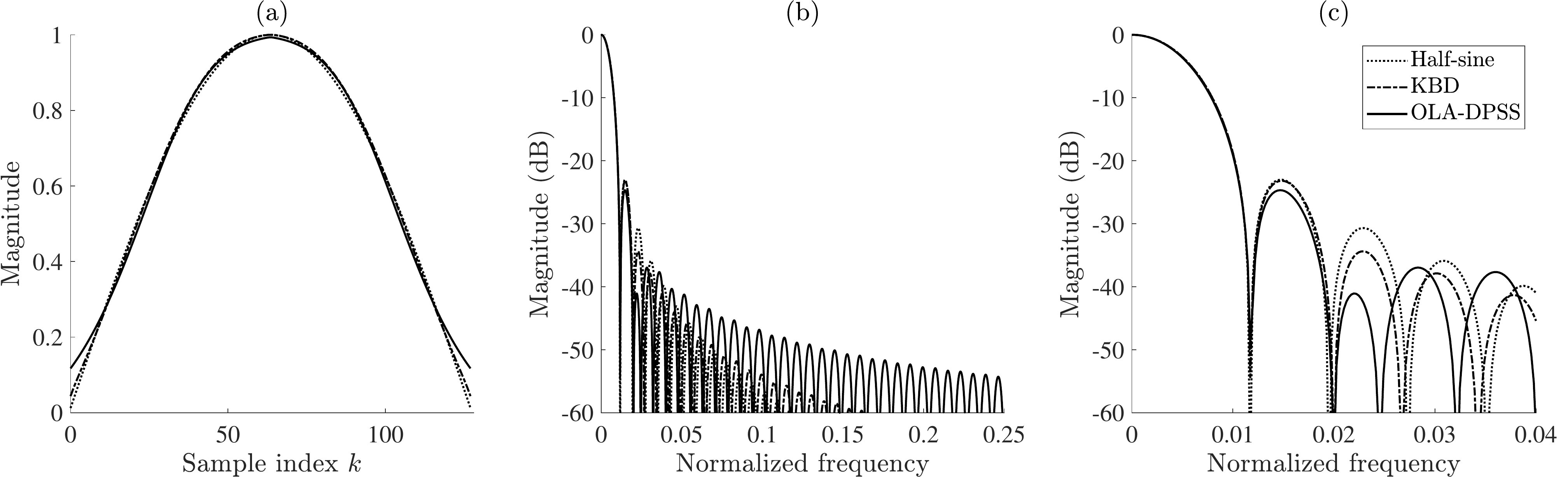}
  \caption{Illustration of (a) the half-sine, KBD and proposed OLA-DPSS windows, (b) their spectral
    responses and (c) responses focused on the central region. Window length is $L=128$ and KBD has $\alpha=4.25$
  and OLA-DPSS has $\alpha=2.75$.}\label{fg:eval1}
\end{figure*}

Figure~\ref{fg:eval1} illustrates the half-sine, KBD and the proposed windows and their spectral responses.
Note that we have here manually tuned the pass-band bandwidth $\alpha$'s in Eqs.~\ref{eq:kaiser} and
Eq.~\ref{eq:toep} such that the main-lobe widths match that of the half-sine window. This choice
allows fair comparison of the side-lobe magnitudes.

\begin{figure*}[t]
  \centering
  \includegraphics[width=0.78\textwidth]{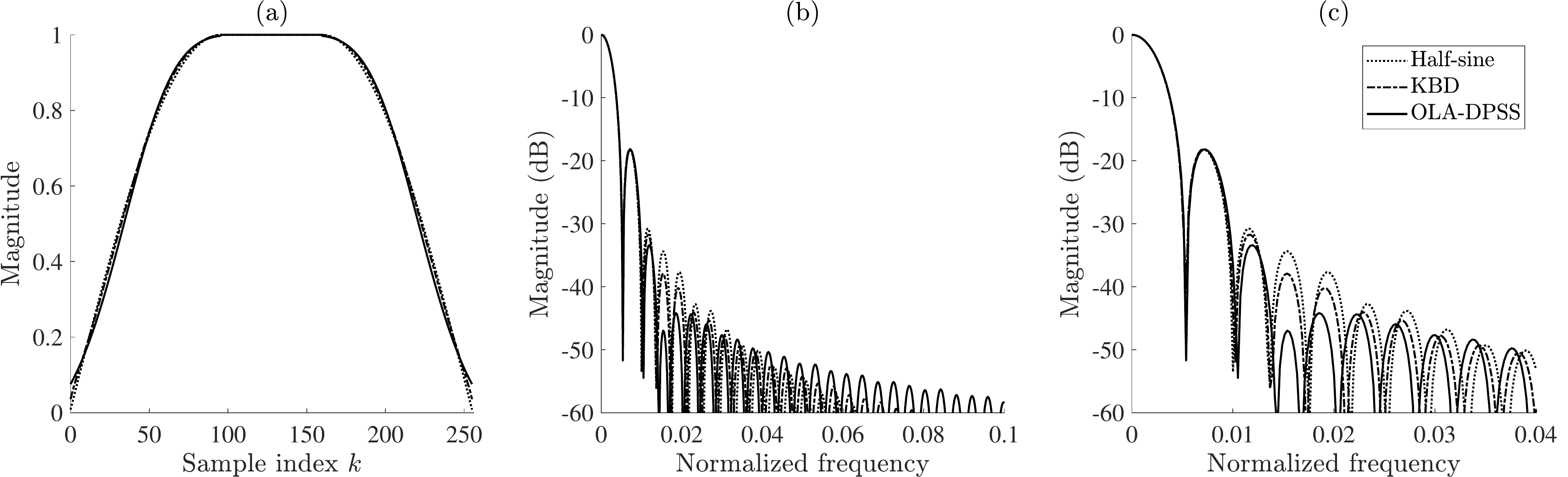}
  \caption{Illustration of low-overlap versions of (a) the half-sine, KBD and proposed OLA-DPSS windows,
    (b) their spectral responses and (c) responses focused on the central region.
    Window length is $L=256$, overlap length is $T=64$, KBD has $\alpha=4.25$
  and OLA-DPSS has $\alpha=5$.}\label{fg:eval2}
\end{figure*}

We observe that the KBD window is in shape very similar to the half-sine, and their spectral responses differ
only for the second side-lobe and higher. The shape of the proposed OLA-DPSS window, however has slightly higher
tails near the ends of the window. Moreover, the spectral response of the OLA-DPSS has an approximately
\SI{2}{\deci\bel} benefit for the first side-lobe. The energy concentration ratios following Eq.~\ref{eq:maxene},
for the half-sine, KBD and OLA-DPSS windows are 16.6559, 16.6582 and \SI{16.6624}{\deci\bel} (parameters
as in Fig.~\ref{fg:eval1}). In other words, by using OLA-DPSS, we obtain
\SI{0.0065}{\deci\bel} and \SI{0.0041}{\deci\bel} improvements in energy concentration in comparison to the
half-sine and KBD windows respectively.

Figure~\ref{fg:eval2} illustrates low-delay versions of the half-sine, KBD and the proposed windows and
their spectral responses. Here we find differences only from the second side-lobe, where the OLA-DPSS is
about \SI{3}{\deci\bel} better than the half-sine and \SI{2}{\deci\bel} better than KBD.
The corresponding energy concentration ratios are 19.6191, 19.6182, \SI{19.6193}{\deci\bel} indicating
that again the OLA-DPSS is the best (by design) but the difference to the others is marginal.

\section{Conclusions}
Design of windowing functions has a long tradition in signal analysis. In processing of speech and audio
signals, we however require that reconstruction of signals is possible. The conventional approach is
to use a method known as overlap-add, where subsequent windows are overlapped such their sum recovers
the original signal. This places constraints
on the window design which has not been adequately taken into account in previous studies.

Slepian windowing functions based on discrete prolate spherical sequences (DPSS) are optimal in terms
of energy concentration, whereby we propose to apply the same objective function but with constraints
that satisfy the symmetries required by overlap-add. The optimization problem is a quadratically constrained
quadratic programming problem, whose solution has become feasible with modern optimization toolboxes.
Since windowing functions are usually determined off-line, computational complexity is not an issue.

The presented evaluations confirm that the proposed overlap-add DPSS or OLA-DPSS windows are efficient
in energy concentration as desired and the proposed window is better than the conventional windows
in all comparisons presented. Since the proposed overlap-add window surpasses the performance of
conventional windows in all aspects, indeed it is the optimal window for this application, OLA-DPSS
should be the preferred choice in speech and audio processing applications.

\bibliographystyle{IEEEbib}
\balance
\bibliography{IEEEabrv,../bibtex_references/refs.bib}

\end{document}